\documentclass[aps,prb,showpacs,twocolumn,floatfix,amsmath,preprintnumbers,nofootinbib,secnumarabic,a4paper]{revtex4}

\usepackage{epsfig,verbatim,hyperref}
\usepackage{amssymb,amsfonts}
\usepackage{graphicx}

\usepackage[caption=false]{subfig}
\usepackage{graphicx}
\def\citen#1{\onlinecite{#1}}

\def\jrn#1#2#3#4#5#6{\textit{#3} \textbf{#4}, #5 (#6).} \def\boo#1#2#3#4#5#6{\textit{#2} (#3, #4, #5).}                           \def\andd{ and } %wspc

\def\boldsymbol#1{#1}
\def\text#1{{\mbox{#1}}}

\def\eqref#1{(\ref{#1})}
\def\eq{Eq.\,}  %epl style

\def\bfl{\begin{flushleft}}
\def\efl{\end{flushleft}}
\def\bfr{\begin{flushright}}
\def\efr{\end{flushright}}
\def\bc{\begin{center}}
\def\ec{\end{center}}
\def\be{\begin{equation}}
\def\ee{\end{equation}}
\def\bse{\begin{subequations}}
\def\ese{\end{subequations}}
\def\ba{\begin{eqnarray}}
\def\ea{\end{eqnarray}}
\def\baa#1{\begin{array}{#1}}
\def\eaa{\end{array}}
\def\bw{\begin{widetext}}
\def\ew{\end{widetext}}

\def\lb#1{\label{#1}}
\def\bit{\begin{itemize}}
\def\eit{\end{itemize}}
\def\bco{\begin{comment}} \def\eco{\end{comment}}
\def\bcs{\begin{cases}}
\def\ecs{\end{cases}}

\def\vena{\boldsymbol{\nabla}}

\def\nc0{\tilde b_0}

\def\vol{{\cal V}}
\def\drm{d}
\def\dvol{\drm\vol}

\def\dn{\rho}  \def\dnc{\bar{\dn}}

\def\upar{\eta}

\begin{document}

\preprint{\small Int. J. Mod. Phys. A \textbf{35}, 2040032 (2020)   
%\ \ [DOI: 10.1209/0295-5075/122/39001]
\quad 
[\href{https://doi.org/10.1142/S0217751X20400321}{DOI: 10.1142/S0217751X20400321}]
}
% https://doi.org/10.1142/S0217751X20400321

\title{
%Logarithmic superfluid theory of physical vacuum: Cosmological singularity problem 
Superfluid vacuum theory
and
%Resolving singularity problem and 
deformed dispersion relations
}

\author{Konstantin G. Zloshchastiev}
%\email{k.g.zloschastiev@gmail.com}

%\email{https://bit.do/kgz}

\affiliation{Institute of Systems Science, Durban University of Technology, 
%P.O. Box 1334, 
Durban 4000, South Africa}

\begin{abstract} 
Using the logarithmic superfluid model of physical vacuum, one can formulate a quantum theory, which successfully recovers Einstein's theory of relativity in low-momenta limit, but otherwise has different foundations and predictions. 
We present an analytical example of the dispersion relation and argue
that it should have a Landau ``roton'' form which ensures the suppression of dissipative fluctuations.  
We show that at small momenta, a dispersion relation becomes relativistic with 
small deformations,
such that a photon acquires effective mass, but a much more complex picture arises at large momenta.
%\keywords{Superfluid vacuum; logarithmic superfluid; photon dispersion relation.}
\end{abstract}

\date{received: 19 Sep 2019 [WSPC]; published: 31 Jan 2019 [IJMPA]}%, xx Sep 2017 [arXiv]}
%\date\today

\pacs{03.75.Kk, 47.37.+q, 47.55.nb, 64.70.Tg\\
%64.70.F− Liquid-vapor transitions
%64.70.Tg Quantum phase transitions
%64.60.Ej Studies/theory of phase transitions of specific substances
%03.75.Kk Dynamic properties of condensates; collective and hydrodynamic excitations, superfluid flow
%47.37.+q Hydrodynamic aspects of superfluidity; quantum fluids
%91.40.-k	Volcanology 91.40.Ft – Eruption mechanisms 47.55.nb	Capillary and thermocapillary flows 47.10.-g	General theory in fluid dynamics 47.20.-k – Flow instabilities 47.35.Fg – Solitary waves
\\ \textbf{Keywords}: superfluid vacuum, logarithmic superfluid, photon dispersion relation
}

\maketitle

%\scn{Introduction}{s:in}
\noindent
It is a general consensus that physical vacuum
%, or non-removable background,
is a nontrivial dynamical matter underlying all phenomena. 
Its dynamics and structure are a subject of intensive studies and debates based on different views and approaches,
which generally agree on a main paradigm of physical vacuum being a quantum object,\cite{dir51} 
but differ in details.\cite{volbook,huabook}
One of such views is superfluid vacuum theory (SVT) --
a post-relativistic approach in high-energy physics and gravity,
which advocates that physical vacuum is a superfluid
and elementary particles are excitations thereof.
% above its ground state.
%; the latter is assumed to be a Bose-Einstein condensate of some kind.
The term `post-relativistic' implies that SVT can generally be a non-relativistic theory,
 but contains relativity 
as a subset, special case or limit with respect to some dynamical value, 
thus fulfilling the correspondence principle.
The superfluid itself is usually understood as a non-relativistic quantum liquid with suppressed 
dissipative fluctuations, thus resulting in zero macroscopic viscosity.\cite{z12eb}
Together with wavefunction's isotropy, this would
make such a vacuum insensitive to the Michelson-Morley-type experiments,
thus different from the classical aether which theory was abandoned long ago.

%\scn{Spacetime induced by background superfluid}{s:me}
%

One can describe quantum liquid by a condensate wavefunction $\Psi$ obeying
a nonlinear wave equation, which can be chosen to have a minimal
 $U(1)$-symmetric form:
$ %\ba
i \partial_t \Psi
=
\left[-\frac{\hbar}{2 m} \vena^2
+
V_\text{ext} (\textbf{x},t)
- 
F (|\Psi|^{2})
\right]\Psi
$, %\label{e:oF}\ea
where $m$ is the mass of a constituent particle,
$F (\rho)$ is a differentiable function on a positive semi-axis $\rho$,
and
%whose first derivative $F' (\rho)$ is integrable on that semi-axis with weight $\rho$.
$V_\text{ext} (\textbf{x},t)$ is an external potential
representing a trapping potential or container (we shall neglect it in what follows).
The condensate wavefunction must be normalized:
$ %\be\lb{e:norm}
\int_\vol |\Psi|^2 \dvol  = 
\int_\vol \dn\, \dvol = {\cal M}
> 0
$, %\ee 
where ${\cal M}$ and $\vol$ are the total mass and volume of the liquid.

SVT assumes that physical vacuum is described by a similar equation, while
photon-like excitations are analogous to  
acoustic waves in superfluid which propagate with the velocity $ c_s \propto \sqrt{p' (\dn)}$,
where prime denotes a derivative.
The correspondence principle requires that in low-momenta limit, SVT must 
recover Einstein's theory of relativity. 
One of postulates of the latter
implies that the speed of photon-like excitations of vacuum should not depend on density,
at least in a leading order with respect to $\hbar$. %a Planck constant.
At low momenta, 
$ c_s $ tends to $ c_0 \approx c $,
where  
$c = 2.9979 \times 10^{10}$ $\text{cm}\, \text{s}^{-1}$
is 
%a universal constant, which is 
historically called the \textit{speed of light in vacuum}.
As shown in Ref. \citen{z11appb},
this results in the following  equation
$ %\be\lb{e:Feq}
\dn
|F' (\dn)|
=
%\mp
m c_s^2/\hbar
\approx
\text{const} (\dn)
$, %\ee
where 
%$ c_s $ is a velocity of photon-like vacuum excitations (SVT analogues of sound waves), and
$\text{const} (\dn)$ is a function which does not dependent on density.
%an approximation sign indicates a leading-order approximation with respect to $\hbar$.
%in our context, $c$ is the speed of the low-momentum massless fluctuations of the physical vacuum, which manifest themselves as photons to a relativistic observer.
The solution of 
this differential equation
%\eq \eqref{e:Feq} 
is a logarithmic function:
$ %\be\lb{e:oFln}
F (\dn) = b \ln{(\dn/\dnc)}
$, %\ee
where $b$ and $\dnc$ are real parameters.
% of the theory.
%; $b$ is also called a nonlinear coupling.
The wave equation thus becomes
\be\label{e:oF}
i \partial_t \Psi
=
\left[-\frac{\hbar}{2 m} \vena^2
+
V_\text{ext} (\textbf{x},t)
- 
b \ln{(|\Psi|^{2}/\dnc)}
%F (|\Psi|^{2})
\right]\Psi
.
\ee
This introduces 
a model with logarithmic nonlinearity
as a further development and generalization of
the non-perturbative theory of quantum gravity with non-exact Lorentz symmetry. %\cite{z10gc,z11pla} 
It tremendously facilitates analytical studies that
the ground state solution for positive values of $b$ was known,
though for purposes other than ours,
since the works 
of Rosen and Bialynicki-Birula and Mycielski.\cite{ros68,bb76}
In Cartesian coordinates,
it is a  Gaussian wave packet modulated by the de Broglie plane wave;
in a rotationally symmetric 3D case, it is a Gaussian function 
of a radius-vector.

%\scn{Dispersion relations}{s:rs}
%

A theory of superfluid vacuum can be formulated in different ways.
The first way is to treat vacuum effects as a small perturbation of a Lorentz-covariant theory.\cite{z10gc,z11pla}
The second, more recent, approach
is to consider particle-like excitations within
the framework of a post-relativistic superfluid vacuum theory where spacetime is an induced phenomenon, as discussed above.
Correspondingly, dispersion relations turn out to be different in each approach.
For the former approach, the relations were derived in Sec. 4 of Ref. \citen{z10gc},
therefore, we do not consider them here; 
we just mention that their energy has a non-polynomial dependence on momentum.
%contrary to other theories with non-exact or broken Lorentz symmetry.
%; though it is still expandable into Taylor series at small momentum or energy.
This alerts us that a perturbative series cannot 
provide a full description of vacuum, therefore the latter requires non-perturbative 
treatment.
%, as much as possible.

\begin{figure}[t]
\centering
\subfloat[$E_p/E_a$]{
  \includegraphics[width=0.99\columnwidth]{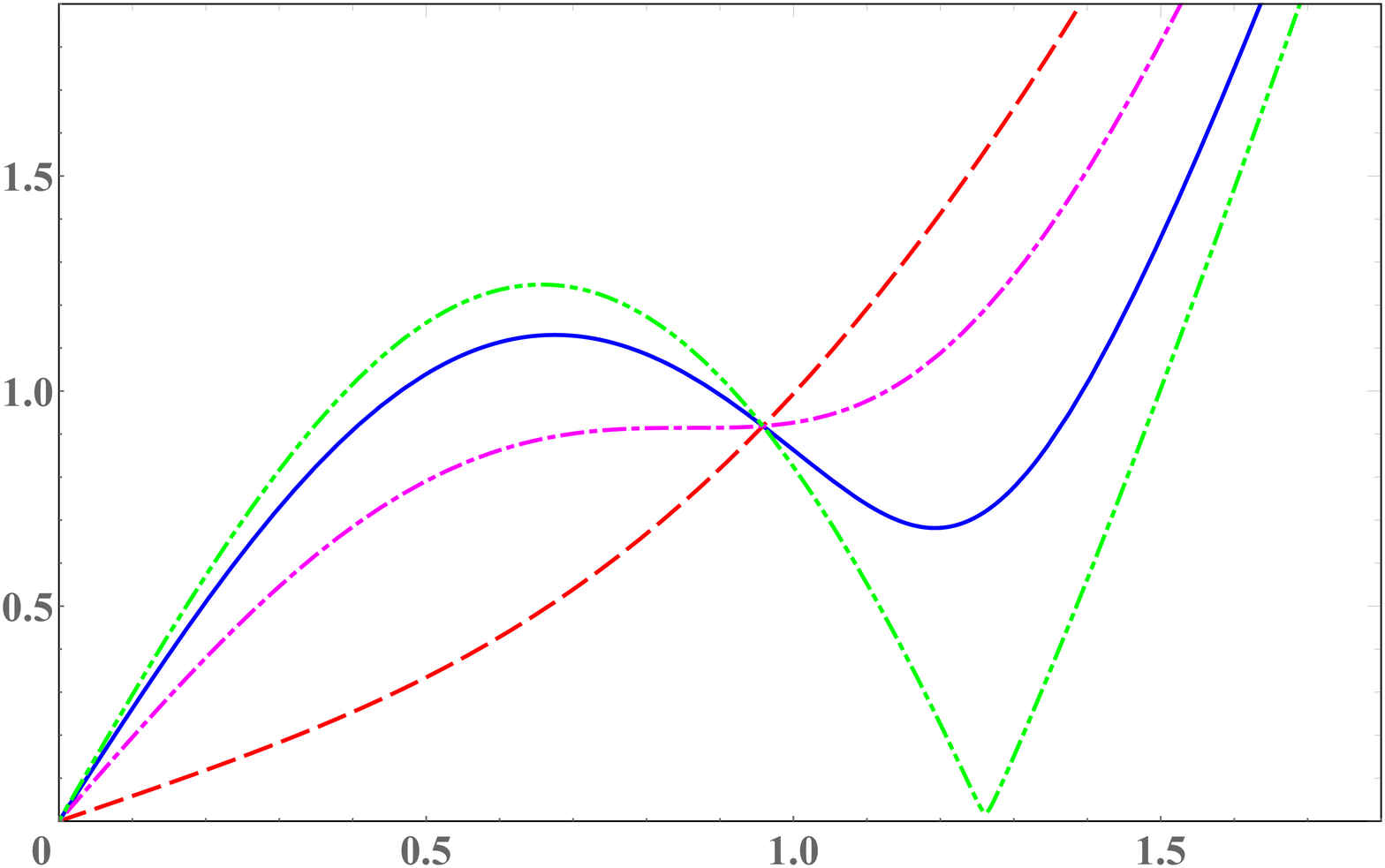}
}
\vspace{3mm}
\subfloat[$c^2_p/c^{2}_0$]{
  \includegraphics[width=0.99\columnwidth]{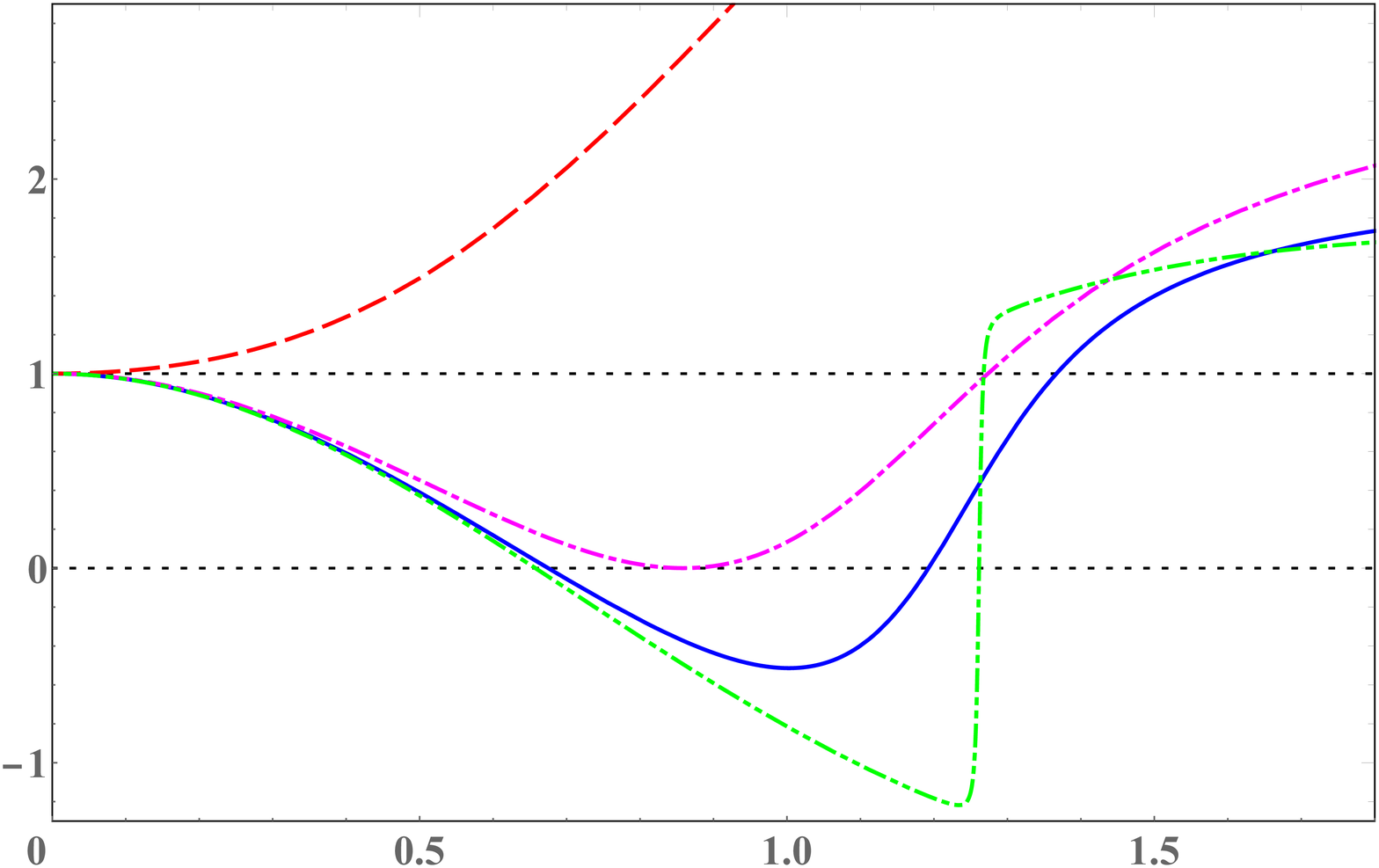}
}
\hspace{0mm}
\caption{Profiles of the energy \eqref{e:disp} and corresponding velocity squared,
% of photon-like excitations of superfluid vacuum, 
versus 
$p /p_a$, for the following values
of $\upar$: 
$1$ (dashed curve), 
$11.43$ (dash-dotted),
$20.66$ (solid),
$25.98$ (dash-double-dotted).
%Negative values of velocity squared indicate an instability region.
}
\label{f:disp}
\end{figure}

The dispersion relations for the second approach 
%are going to be a main focus of our study. They 
can be introduced by analogy with conventional superfluids.
%, because an easiest way to see how does relativity arise in superfluid vacuum theory is the energy spectrum of particle-type excitations of a typical superfluid, such as He II.
In superfluid $^4$He, energy as a function of
momentum has a distinct shape, usually dubbed the Landau ``roton'' form.
As the excitation's momentum increases, this energy grows from zero,
until it reaches a local maximum
(called the \textit{maxon} peak), which is crucial for suppressing dissipative fluctuations.\cite{z12eb}
Then it dives down to a local minimum (called the ``roton'' minimum, for historical reasons),
and eventually it starts growing again, all the way to a border 
of a theory's applicability domain, \textit{akin} to a solid curve in Fig. \ref{f:disp}a.
%\textit{ad infinitum}.  
In the regime of small momenta, 
%called the \textit{phononic} regime,
the dispersion relation is approximately linear with respect
to momentum, which is a behaviour typical for relativistic particles.
%, if one replaces speed of light by a speed of sound.
%As a matter of fact, one can still use some sort of relativistic description (by adding extra fields to account for small deviations from the linear law), until the momentum reaches its value corresponding to the ``maxon'' peak. From there up, relativistic approximation is no longer robust or natural.

We thus 
%project the above-mentioned picture from the condensed-matter realm to the realm of elementary particle physics and gravity (by replacing phonons with photons, speed of sound with speed of light, etc.):
assume that the physical vacuum has a similar energy spectrum of
excitations - with its own values of parameters of course. 
%then one can .
%Moreover, one can further explore this analogy to find the analogs of liquid-helium phenomena in high-energy physics and quantum gravity. 
It also means that one can use the relativistic approach (with perhaps some tweaking by adding additional fields)
for a large range of energies - until momentum reaches a value corresponding to 
the maxon peak,  which can be rather large, up to a Planck scale.
These intermediate models can still provide valuable understanding
about various fundamental phenomena,
such as the mass generation mechanism and non-zero extent of particles.\cite{z11appb,az11,dz11,dmz15}
However, conceptually new physics steps in near and above
the maxon threshold; 
%an essentially non-relativistic quantum regime occurs;
vacuum Cherenkov radiation and 
luminal boom being examples thereof.\cite{z11pla}

An analytical example of a dispersion relation, 
which has a Landau form and correct proportions
%between positions of 
for local extrema, is given by the formula:
\bw
\be\lb{e:disp}
E_p
=
\frac{p^2}{2 M_a}
\sqrt{1 + 
%\frac{1}{2}
\upar\,
 f\!\left(
\frac{p}{p_a} \right)
%}{p^2}
}
,\ \
f (k)
\equiv
\frac{1}{2 k^2}
\left[
1 - 
\aleph\, \text{e}^{-k^2}
+
\left(
\frac{1}{k}
- 2 k
\right)
D_+ (k)
\right]\!,
\ee
\ew
where
$D_+ (k)$ is a Dawson function,
$
\aleph = \sqrt\pi/2 + 1/(\text{e} \, \text{erf}(1)) \approx 1.32
$,
$p_a = 2 \hbar/a$ is de Broglie momentum corresponding
to the length scale
$a = \hbar/\sqrt{2 m |b|}$,
and $M_a = p_a^2/2 E_a
%$ is a ``dressed'' mass of a condensate particle, $M 
\propto m$.
%$\vec k = \vec p/\hbar$ is the wave vector.
Here, $p_a$, 
$E_a$ and $M_a$ set the momentum, energy and mass scales of the theory,
whereas the parameter $\upar$ controls local extrema of the dispersion curve:
one can check that
$E_p$ acquires the Landau form, cf. a solid curve in Fig. \ref{f:disp}a, only if
$11.43 < \upar < 25.98$.

Furthermore, Fig. \ref{f:disp}b displays 
%a profile of
the excitations' velocity $c_p \equiv \sqrt{d E_p/d p}$
divided by its low-momentum value 
$c_0 = \lim\limits_{p\to 0}c_p$. 
As in Fig. \ref{f:disp}a, it is the solid curve which is of interest to us.
One can see that at small momenta, its behaviour 
follows the relativistic pattern, but becomes a rather nontrivial as $p$ grows.
Initially it decreases with increasing $p$, which corresponds to a photon's slow-down
and thus it can be interpreted as a photon acquiring the effective mass
$\mu$ --
when expanded in Taylor series
at small $p$,
\eq \eqref{e:disp} expectantly yields 
a deformed-relativistic dispersion:
\bw
\be\lb{e:disp1}
E_p^2=
c_0^2 p^2 
+ \mu^2 (p)\, c_0^4
, \ \
\mu (p)
\equiv
\left\{
A_4 (p/p_a)^4
+
A_6 (p/p_a)^6
+
{\cal O}\!\left[ (p/p_a)^8 \right]
\right\}^{1/2}
,
\ee
\ew
where 
$c_0 = (p_a/ 2 M_a) \sqrt{\upar \, (1 - \aleph/2)}
$,
$A_4 
%=  (1 - \upar/3) E_a^2/c_0^4 - p_a^2/c_0^2
=
(p_a/c_0)^2 (\aleph - 8/3 + 2/\upar)/(2 - \aleph)
$,
$A_6 
%= (3 \upar/10) E_a^2/c_0^4 + (1/2) p_a^2/c_0^2
=
(p_a/c_0)^2 (8/5 - \aleph/2)/(2 - \aleph)
$,
and
${\cal O} [k^n]$ denotes terms of order $k^n$ and above.

Note that this effective mass generation mechanism is different from 
the gap mechanism in conventional quantum condensed matter such as superconductors:
it is a post-relativistic effect and 
it happens before one reaches any of the local extrema.
Since $c_0 \approx c$, the following experimental constraint applies
to our parameters:
$a M_a/\sqrt{\upar} = (\hbar/c_0) \sqrt{1- \aleph/2} \approx 
2.04 \times 10^{-38} \, \text{g} \, \text{cm}$,
whereas $A_6$ does not explicitly depend on $\upar$.

A series \eqref{e:disp1} converges only at $p < p_a$:
in the vicinity of local extrema $p \sim p_a$, not to mention at $p\to \infty$, a formula changes drastically,
which reaffirms necessity of non-perturbative treatment.
As momentum grows, energy reaches first the maxon peak which marks the beginning of a region
in which $c_p^2$ becomes negative thus indicating absence of any classical propagation.
%, except perhaps quantum tunneling.
Starting from that region, classical 4D spacetime is no longer a robust description,
and the definition of a particle can no longer be based on irreducible representations of a Poincar\`e group.
This region extends until object's momentum approaches a value corresponding to the ``roton'' minimum, 
after which the maximum attainable velocity can eventually go above $c_0$.
% as $p$ grows. 
This is where the luminal boom occurs,
which is the vacuum analogue of sonic boom in air.\cite{z11pla}
Beyond that value, 
breakdown of the quantum liquid occurs:
a moving object or observer no longer experiences vacuum condensate nor induced spacetime.
%The latter phenomenon is somewhat similar to what happens to conventional Bose-Einsteing condensates when temperature rises. 

%\scn{Conclusion}{s:si}
%

%\section*{Acknowledgments}
%\bc *** \ec
%~\\ \textbf{Acknowledgments}\\
\begin{acknowledgments}
%I am grateful to participants of the 8th International Conference on Applied Physics and Mathematics ICAPM-2018 in Phuket, Thailand (27-29 January 2018) \cite{z18cs1}, and XXIII Fluid Mechanics Conference KKMP-2018 in Zawiercie, Poland (9-12 September 2018) \cite{z18cs2}, where parts of this work were discussed. 
% as well.
This work is 
%based on the research 
supported 
by Department of Higher Education and Training of South Africa and 
in part by National Research Foundation of South Africa. 
%under Grants Nos.95965 
%98083 and 98892,
%and CMC Department of People's Republic of China.
%This work is based on the research supported wholly / in part by the National Research Foundation of South Africa (Grant Numbers xxx, yyy, and zzz)
Proofreading of the manuscript by P. Stannard is greatly appreciated.

\end{acknowledgments}

%\appendix

%\section*{References}

\end{document}